\documentclass[11pt]{article}
\usepackage[utf8]{inputenc}
\usepackage[T1]{fontenc}
\usepackage{graphicx}
\usepackage{longtable}
\usepackage{wrapfig}
\usepackage{rotating}
\usepackage[normalem]{ulem}
\usepackage{amsmath}
\usepackage{amssymb}
\usepackage{capt-of}
\usepackage{hyperref}
\usepackage{setspace}

\usepackage{mathtools}
\usepackage{microtype}
\allowdisplaybreaks

\usepackage[
  backend=biber,
  style=alphabetic,
  sorting=ynt
]{biblatex}

\DeclareUnicodeCharacter{2212}{-}
\author{Chris Scott}
\date{\today}
\hypersetup{
 pdfauthor={Chris Scott},
 pdftitle={Turbulence from First Principles},
 pdfkeywords={turbulence, energy cascade, ergodic solution, de Sitter, two-potential formalism, dyality},
 pdfsubject={},
 pdfcreator={}, 
 pdflang={English}}

\newcommand{\pt}{\partial_t}

\newcommand{\lap}{\nabla^2}

\newcommand{\gr}{\varphi}

\begin{document}
\title{Turbulence from First Principles}
\maketitle

\begin{abstract}
We develop a first-principles route to turbulence by using electrodynamics of continuous media in the viscous limit to obtain a Navier--Stokes--type model under a relaxed long-wavelength approximation (LWA). We treat oscillators with two orthogonal angular momenta as a spin network with properties applicable to the Kolmogorov--Arnold--Moser (KAM) theorem. The microscopic viscous limit has an irreducible representation that includes $O(3)$ expansion terms for a radiation-dominated fluid with a Friedmann--Lema\^\i tre--Robertson--Walker (FLRW) metric, equivalent to an oriented toroidal de Sitter space. The combined particle-velocity and pressure-flux is conserved on the hypergeometric $n$-tori with stochastic boundary conditions. Under the relaxed LWA the turbulence solution in $\mathbb{R}^{3,1}$ lies on $\binom{6}{3}$ de Sitter intersections of three orthogonal $n$-tori.
\end{abstract}

Energy cascade describes the process of momentum transfer from large fluid rotations into ever smaller scale rotations. Kolmogorov scaling applies to the inertial sub-range of the power spectrum and provides a universal scaling law in which the energy contained within the domain of the eddy is proportional to its characteristic length scale $l$ and its wave number $k=l^{-1}$. The characteristic energy of the eddy ensemble $E$ is $E(k) = \alpha \varepsilon^{2/3} k^{-5/3}$, where $\varepsilon$ is the ensemble energy dissipation rate and $\alpha$ is a constant \cite{ortizsuslow_evaluation_2019}. In his seminal 1941 statistical approach Kolmogorov assumed local isotropy of fluid pulsations in a medium equipped with the \textit{homogeneity} property \cite{kolmogorov_local_1991}, and this property being invariant under reflection and rotation transformations. In that analysis the velocity field is expanded to second order, and the transformations are evaluated at a high-frequency, homogeneous boundary condition.

\begin{figure}
    \centering
    \includegraphics[width=1\linewidth]{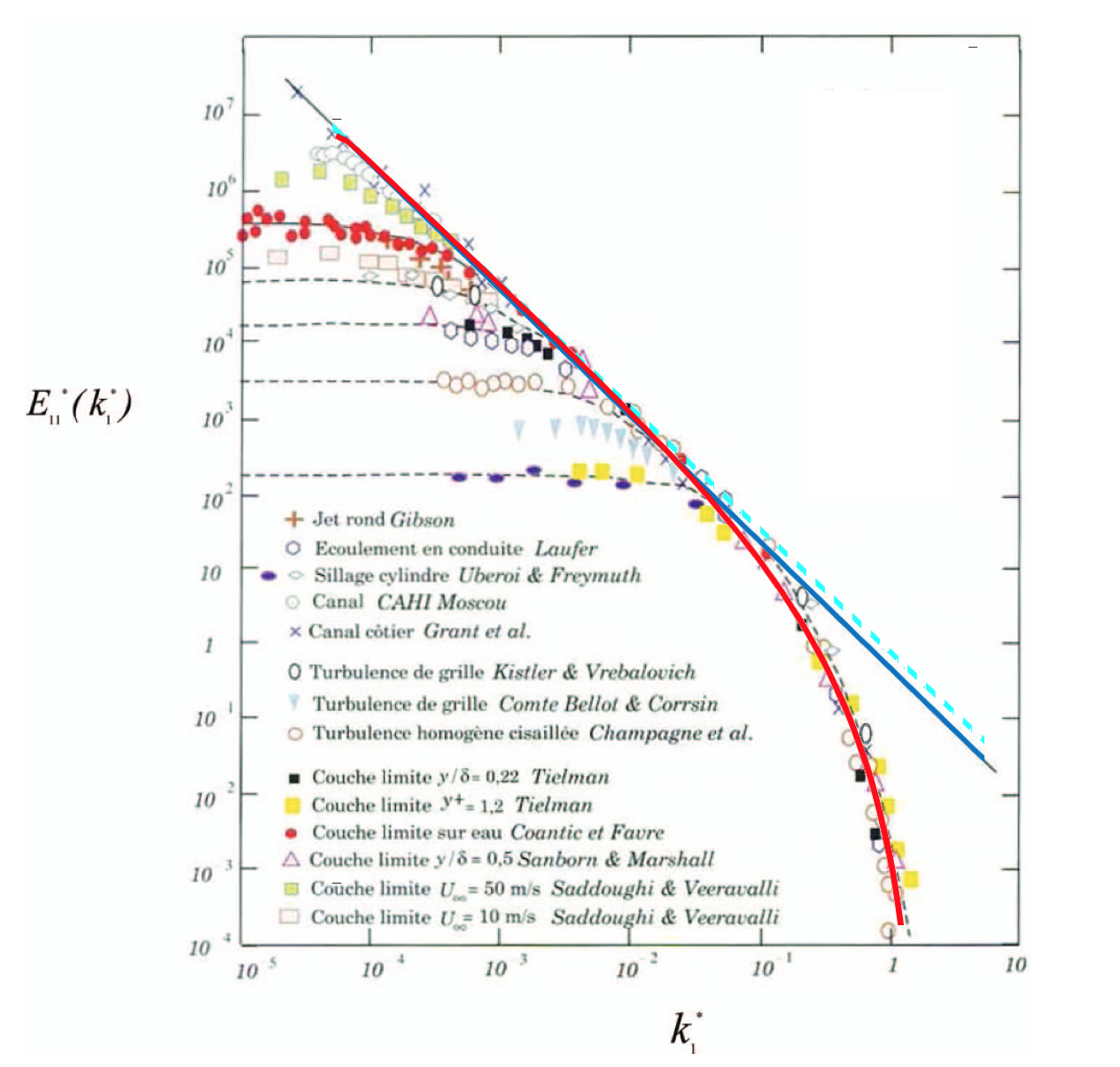}
   \caption{Comparison of our analytic result (red line) to empirical energy cascade in turbulence. Power spectrum of a fluid exhibits Kolmogorov scaling (blue solid line) in the inertial subrange before viscous effects dominate at high wavenumber $k$. Our result converges on a \emph{half-golden-ratio} inertial subrange exponent (dashed cyan), i.e.\ $(1+\sqrt{5})/4=\gr/2$ with $\gr=(1+\sqrt{5})/2$. $E^*_{11}(k)$ is the normalised longitudinal energy spectrum of the eddy domain ensemble with normalized wavevector $k^*$. The viscous scaling solution was evaluated using higher order terms of the velocity field to $O(3)$. One-dimensional turbulence spectrum data for various Taylor microscale Reynolds numbers with a fitted (dashed) curve adapted from \cite{chassaing_turbulence_2000, antonia_collapse_2014}.}
    \label{fig:k-scaling}
\end{figure}

\begin{figure}
    \centering
    \includegraphics[width=0.75\linewidth]{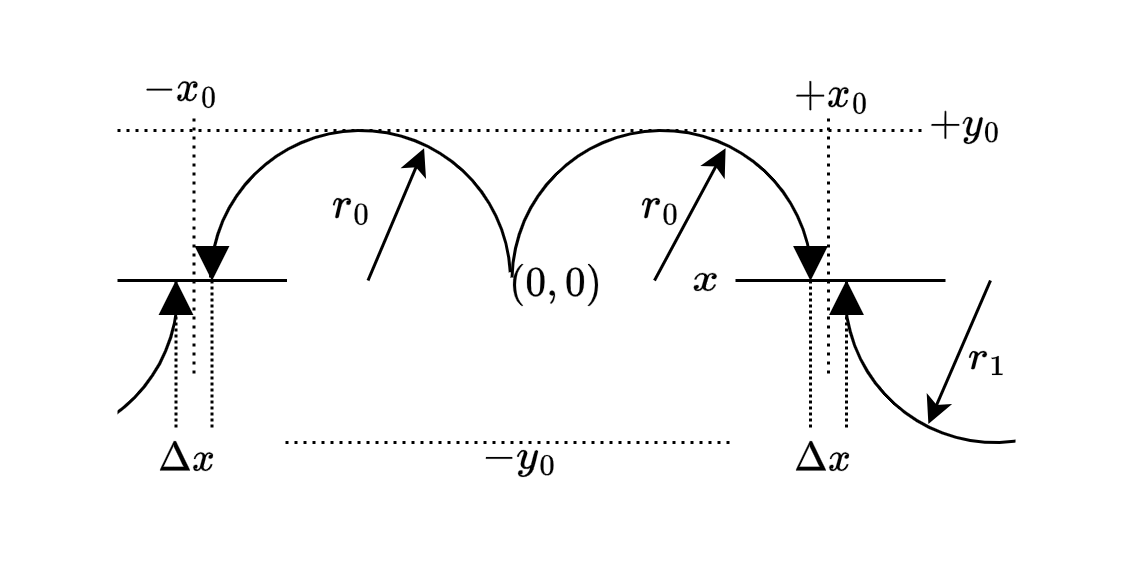}
    \caption{For paths traced by the electric displacement vector $\vec{P}$ the maximum dissipation via viscous effects occurs in the limit $\Delta x \rightarrow r_{\text{bound}}$ and $r_0=r_1=k_{\max}^{-1}$. Toroidal moments arise from octupole ($O(3)$) irreducible terms and higher as in the configuration shown, having four dipoles centred about $\pm x_0$. They equally arise from the in-plane rotation of quadrupole moments, that is \eqref{eq:rofstrain} in a rotating frame. Angular momentum is transported via \eqref{eq:rofstrain} through either the $\pm y_0$ surface ($\parallel T^{(m)}$), or the $\pm x_0$ surface ($\perp T^{(m)}$).}
    \label{fig:retard0}
\end{figure}

\paragraph{The motivation.} Including third-order expansion terms of the velocity field allows for the assessment of turbulence within the framework of the complete set of polarizations, applicable to any continuous media \cite{dubovik_toroid_1990,dubovik_material_2000, papasimakis_electromagnetic_2016}. We may gain insight into the reverse energy cascade process in which self-organisation takes place and identify phase transition mechanisms of \textit{any} medium in motion provided that Maxwell's equations are relevant at some scale. The phenomenon of \textit{intermittency} at high wavenumber is closely related and may be assessed using a multi-fractal framework despite the fact that a first-principles solution has remained elusive \cite{salazar_stochastic_2010, poursina_stochastic_nodate}. Similarly, the shell model of magnetohydrodynamic turbulence has enjoyed some success in 2D and limited 3D regimes \cite{plunian_shell_2013} by employing a quasi-condensed-matter approach with, however, some paradoxical results.\footnote{\cite{plunian_shell_2013} cites nonphysical energy conservation due to helicity of the magnetic field. This may be explained by the oft-overlooked toroid polarization as it captures curl of the magnetic and electric total flux, and specifically treats the contraction of magnetic dipoles to effective magnetic monopoles which has historically been a seed for confusion.} Recent stochastic approaches to model 1D turbulence with small viscosity appear successful \cite{kuksin_kolmogorovs_2022} and consistent with \cite{kolmogorov_local_1991} but fail with the introduction of viscosity. We approach the problem from the high viscosity, heat-death limit by employing the toroid polarization of the Lorentz invariant two-potential formalism. At high frequency, the $O(2)$ truncation resolves only linear pulsations of the velocity field, whereas the $O(3)$ expansion retains the resolution-limit rotations and thereby incorporates viscous effects and their associated dynamics. Throughout we work under a relaxed LWA, which retains near-field multipoles (including toroidal polarisation) and thereby necessitates the dual-potential formulation.
\\
Instead of using the K41 assumption of \textit{isotropy} we evaluate the angular momentum dispersion ratio $\frac{\dot{q}_{\parallel}}{\dot{q}_{\perp}}\big|_{k_{\max}}$ assuming \textit{degeneracy} of the classical motion of charged particles' oscillation. Specifically, $r_{\min}$ in this limit is the smallest accessible radius traced by a bound-charge electric displacement vector $\vec{P}$ where the total electric flux displacement vector is $\vec{D}$, following the convention of \cite{dubovik_material_2000}, 
\begin{align*}
    \vec{D} = \vec{E} + 4\pi \vec{P}.
\end{align*}
The microscopic electric field flux vector $\vec{E}$ contributes to macroscopic pressure once integrated over solid angle (schematically over $4\pi$) in kinetic/transport treatments \cite{braginskii_transport_1965}. Similarly, viscosity is the macroscopic property directly related to the rate of strain tensor,
\begin{equation}
\label{eq:rofstrain}
    D_{ij}^{(2)} = \int{\big(\vec{r}_{i}\vec{P}_{j} + \vec{r}_{j}\vec{P}_{i} - \tfrac{2}{3} (\vec{r}\cdot\vec{P}) \delta_{ij}\big)}\,d\vec{r}
\end{equation}
where $\vec{r}_x = \Delta x$ as shown in Fig.~\ref{fig:retard0}. \\

Let a unit cylinder have radius $R+r$ and height $2r$ with an embedded torus with minor and major radius $r_0$ and $R$ respectively. The configuration for maximum shear strain evaluated at the cylinder surface defined by a periodic surface $x_0, y_0$ is shown in Fig.~\ref{fig:retard0}. The volumetric toroid moment for a purely poloidal current confined to the torus surface is
\begin{align}
    \frac{\vec{T}^{(m)}}{V} = \frac{|\vec{P}|\,\pi R r}{c}.
\end{align}
For a fixed poloidal current magnitude we can achieve maximum viscosity $\int_G |D_{ij}^{(2)}|\, d^3r$ in the domain $G \in \mathbb{T}^2$ for the irreducible configuration shown in Fig.~\ref{fig:retard0}. For a system representation containing only angular momentum the spin networks closed in $\mathbb{R}^3$ have a characteristic ensemble energy $\oint{{D}^{(2)}_{ij \parallel}}\,dl$ such that flux field lines are always closed at some scale. The picture is analogous to a spin network \cite{rovelli_covariant_2014} where angular momentum packets are irreducible to a choice of spin-$\tfrac{1}{2}$ or spin-0 on a discrete spacetime. $\mathbb{T}^2$ is a natural coordinate frame to start in to describe the oriented ($\hat{x},\hat{y},\hat{z}$) viscous dissipation properties. 

\textbf{From KAM theory:} Implying some characteristics of invariant tori from the KAM theorem, the stability of orbits under perturbation leads to a dispersion ratio $\dot{D}^{(2)}_{ij \parallel} / \dot{D}^{(2)}_{ij \perp}$ being half the winding number of the most statistically observable toroidal orbit, $\frac{k_{\parallel}}{2k_{\perp}}$. The rate at which degeneracy is reached is proportional to the irrationality of the winding number, the golden ratio being the most irrational \cite{irwin_quantum_nodate}. We can use the concept of nearest-neighbour tori in a periodic packing to assess the momentum transfer in analogy to the multi-fractal approach \cite{benzi_multifractal_2022}. Longitudinal momentum transfer can occur only via \eqref{eq:rofstrain} between stacked tori, whereas half of the transverse contact area is shielded due to the inward converging geometry of the horn torus, $\frac{r}{r+R} \rightarrow \frac{1}{2}$ when $n\rightarrow\infty$. For a fully developed, degenerate system of orbits at $k_{\max}$,
\begin{align} \label{eq:dispersn}
    \frac{\dot{D}^{(2)}_{ij \parallel}}{\dot{D}^{(2)}_{ij \perp}} = \frac{1+\sqrt{5}}{4} = \frac{\gr}{2}, \qquad \gr:=\frac{1+\sqrt{5}}{2}.
\end{align}

\begin{figure}
    \centering
    \includegraphics[width=1\linewidth]{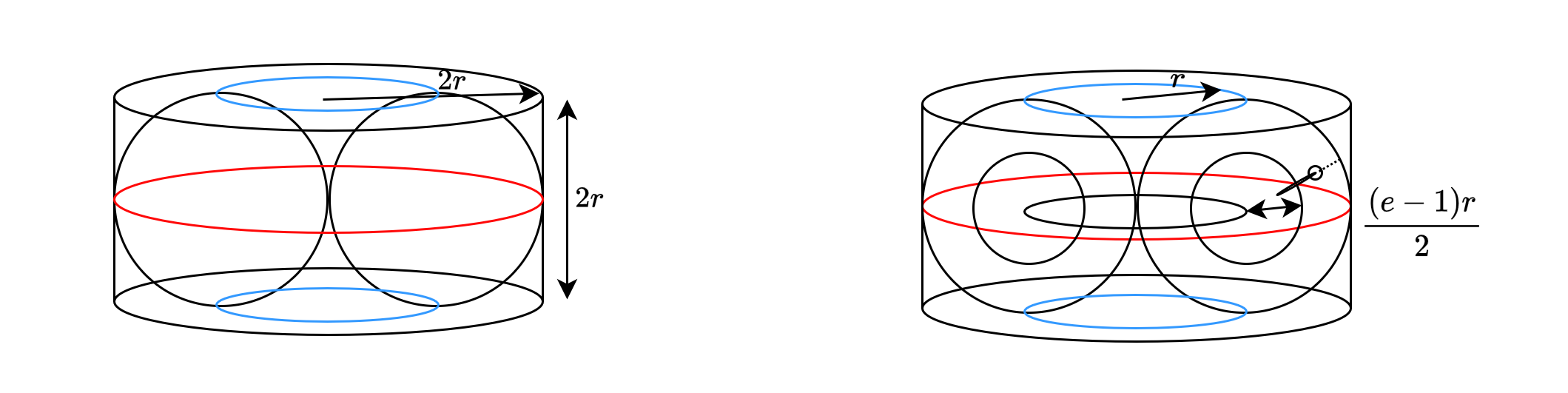}
    \caption{Left, unit cylinder with embedded torus showing longitudinal contact sites (blue). Right, fractally enveloping ring tori converge onto a dense $n$-dimensional horn torus filling the unit cylinder when $r_{n+1} = \frac{r_n(e-1)}{2}$, $n\rightarrow \infty$.}
    \label{fig:unit_cyl}
\end{figure}

The invariant rotation of EM fields and sources is determined by a self-consistent formalism \cite{dubovik_material_2000}. A $U(1)$ \textit{dyality}\footnote{the rotation on EM fields and EM sources} rotation captures the scale dependence introduced by closed electromagnetic field lines \cite{han_manifest_1971, bukina_higher_2001}. The domain $G$ may be constructed with fractally enveloping tori $k_-$ and fractally contained tori $k_+$ up to $k_{\max}$ for a single particle with two orthogonal angular momenta $k_n, k_{n+1}$. The Fourier-containing chart for any path interval $\int{\vec{D}\,dt}$ in the domain $G$ onto $\mathbb{R}^3$ happens to correspond exactly with the FLRW-based toroidal de Sitter flat chart \cite{numasawa_global_2019},
\begin{align} \label{eq:ds}
    ds^2 = -dt^2 + e^{2Rt}d\vec{R}^2 ,
\end{align}
where the scaling constant $R$ is proportional to the radius of the particle (ensemble) path $r$ on the oriented $\mathbb{T}^2$ space with coordinates $\vec{R}$. The dyality relation gives the $n$th-order multipole of the fractally enveloping $n_-$, and containing $n_+$ toroids. 
\begin{align}
\label{eq:Rscalar}
{R}_j
&=
\frac{2}{(e-1)} \frac{(n_j-n_{\min})!}{k_j - k_{\min}}
-
\frac{e-1}{2\,(n_{\max}- n_j)!\,(k_{\max}-k_j)},
\qquad n \in \mathbb{Z}^+,\; k \in (0,1]
\\
\label{eq:dyality1}
\vec{T}_{z}^{n(m),(e)}
&=
\vec{T}_{z}^{n+1(e),(m)}
=
\nabla_z \times {\vec{T}^{n-1(e),(m)}}
\\
\label{eq:dyality2}
\nabla \times \vec{T}^{(e),(m)}
&=
\pm \Big( \frac{1}{c} \frac{\partial \vec{T}^{(m),(e)}}{\partial t}
+ \nabla \phi^{(m),(e)} \Big)
\end{align}

Superscript $(m)$ denotes the magnetic toroidal moment and $(e)$ denotes the electric toroid moment \cite{dubovik_toroid_1990}. The spatio-temporal interchange between closed magnetic and electric field lines and between hypersurfaces is evident in \eqref{eq:dyality1} and \eqref{eq:dyality2}.

\paragraph{Applicability to neutral media.}
By “neutral” we mean flows with negligible free charge density and small bulk polarizability at the hydrodynamic scales of interest. Our construction targets media where a finite microscopic susceptibility renders near-field electromagnetic effects non-negligible (for example, polar dielectrics, plasmas, or composites with mesoscale structure). In strictly neutral, weakly polarizable fluids (air, water at standard conditions) the present closure should be interpreted as an effective model, and the mapping between microscopic Coulomb interactions and macroscopic pressure must be made explicit (for example, via Braginskii-type relations \cite{braginskii_transport_1965}).

The initial position, momentum and pressure flux of the $j$th particle (or ensemble of particles) is defined on a single hypersurface $n_j$, in toroidal coordinates,
\begin{align} \label{eq:position}
    \vec{R}_{j0}^2 = e^{\phi_0}\big((e-1)R_0\big)^2 + \big( e^{\phi_1 i}R_1 + \cdots +e^{\phi_{n_j} i}R_{n_j}\big)^2 .
\end{align}
All-order toroidal polarized electromagnetic interactions may be solved self-consistently on $\binom{6}{3}$ intersections of three orthogonal FLRW (toroidal de Sitter) spaces corresponding to a radiation-dominated fluid at the microscopic boundary condition of viscous heat-death. The scaling parameters are derived from $k_{\max}$ and $n_{\max}$ as properties of the medium, $k_{\min}$ and $n_{\min}$ as properties of the domain $G$ and $n_j,k_j$ as the properties of the particle (or ensemble of particles) under consideration. The total flux displacement $\vec{D}$ is conserved on these intersections with emergent stochastic properties. 

\paragraph{Ansatz.} We seek solutions at the stochastic intersections of three orthogonal toroidal de Sitter spaces described by \eqref{eq:ds}, \eqref{eq:Rscalar}, \eqref{eq:dyality1}, \eqref{eq:dyality2} with the dispersion relation \eqref{eq:dispersn} of the degenerate spin-oscillator system. 

\paragraph{Derivation.} The toroidal momentum de Sitter space flat chart metric is \eqref{eq:ds}. The oriented toroidal coordinate space $\mathbb{T}^2$ maps to the oriented toroidal phase space $\mathbb{T}^1$ for the fractal 2D-harmonic oscillator set. The Hamiltonian for the motion of a particle on a torus in fractal tori phase space is given by
\[
\Omega = \frac{q_{\phi_n}^2}{2m} + \frac{q_{\phi_{n-1}}^2}{2m},
\]
where $q_{\phi_n}$ and $q_{\phi_{n-1}}$ are the conjugate momenta corresponding to the angular coordinates $\phi_n$ and $\phi_{n-1}$ respectively to form a torus, and $m$ is the mass of the particle containing the bound charge. The ray equations for this spin-orbit system correspond to the free motion of charged particles on tori in coordinate space,
\begin{align*}
    \frac{d\phi_n}{dt} = \frac{q_{\phi_n}}{m},\quad
\frac{d\phi_{n-1}}{dt} = \frac{q_{\phi_{n-1}}}{m}, \quad
\frac{dq_{\phi_n}}{dt} = 0,\quad
\frac{dq_{\phi_{n-1}}}{dt} = 0.
\end{align*}

Any particle in $\mathbb{R}^3$ lives on the $n$th toroidal hypersurface given a momentum and position \eqref{eq:position}. The electromagnetic scaling inter-dependency of \eqref{eq:dyality1} and \eqref{eq:dyality2} is made explicit with the fractally enveloping torus minor and major radii, 
\begin{align} \label{eq:rtoR}
    r_n = k_n^{-1} = \frac{R_n(e-1)}{2},
\end{align}
so that the entire space (of a 2D spin-orbit system) can be described by two independent complex constants of motion,\footnote{For the degenerate case there is only one locally defined integral of motion: $\mathrm{Im}\big(\log(R_n)\tfrac{2k_n}{e-1} - \log(r_n)k_n\big)$.} \cite{jose_classical_2006} 
\begin{align}
    F_{(n-1)} = R_n e^{i 2k_n t/(e-1)} , \quad F_{n} = r_{n} e^{i k_n t}.
\end{align}
For invariant tori in phase space a 2D oscillator has incommensurate frequencies $\tfrac{\omega_{k+1}}{\omega_k}$ corresponding to an irrational winding number and is most robust to perturbation. The resulting dispersion ratio is \eqref{eq:dispersn}.  
\\[4pt]
The microscopic electric field flux $\vec{E}$ is used here as a proxy for the microscopic pressure-gradient contribution to momentum flux (distinct from macroscopic Ohmic-response terminology), and should not be confused with the electric bound-charge displacement vector $\vec{P}$. The total displacement flux is
\begin{align} \label{eq:dep_pressure}
    \vec{D} = \vec{E} + 4\pi \vec{P} = \vec{p} + 4\pi \vec{u}\,dt.
\end{align}

We now introduce a component-wise, dimensionless rate ratio that will be used to normalise the higher-order (primed) contribution:
\begin{align}
\label{eq:GammaDef}
\Gamma_j \;:=\;\frac{\pt D'_j}{\pt D_j}, \qquad j\in\{x,y,z\}.
\end{align}

The current configuration at the viscous heat-death boundary condition is defined as the microscopic magnetic dipole moment $\vec{M}_z \rightarrow 0$ \cite{braginskii_transport_1965}, evaluated at $n_{\max}, r_{\min}$. In Cartesian coordinates and in the convention of \cite{nemkov_electromagnetic_2018},
\begin{multline}
 \label{eq:nemkov1}
    \frac{\partial \vec{D}'(\vec{R}(n_{\max}-1))}{\partial t} = 2\pi^2 R \frac{\partial\vec{D}(\vec{R}(n_{\max}))}{\partial t} \nabla \times \hat{\mathbf{z}}\, \delta(\vec{r}) \\ + \frac{\pi^2 R}{8} \frac{\partial\vec{D}(\vec{R}(n_{\max}))}{\partial t} \left[2R^2 \big(\nabla^2_x + \nabla^2_y\big)\right] \nabla \times \hat{\mathbf{z}}\, \delta(\vec{r}) + O(r^{-4}) .
\end{multline}
Here $\vec{D}'$ is the total electric flux displacement vector with path curvature $\tfrac{2}{e-1}r_{\min}$, corresponding to the penultimate cascade momentum. Applying \eqref{eq:nemkov1} to currents confined on $n$-tori oriented in the $y$-direction allows us to simplify to current loops on the $x$--$y$ plane given $\vec{M}_z =0$ (see Fig.~\ref{fig:retard0}). We can drop the vector notation since $R$ is a scalar and $c =1$ at the microscopic viscous limit scale $r_{\min}$ with a guiding centre radius defined by \eqref{eq:Rscalar}. Since the irreducible representation of all three polarization states $(E,B,T^{(m),(e)})$ is two-dimensional we can represent $J_{\mathrm{tot}}$ as the total current inscribed along the $n$th meridian path of an $n$-torus. The single ray then is confined on a circle with radius $R$. Continuing the example of Fig.~\ref{fig:retard0}: 
\begin{align} \label{eq:current}
   \vec{S}_n J_{\mathrm{tot}} = \vec{S}_n\frac{\partial D}{\partial t}, \quad \vec{S}_n(\phi) = \frac{d}{d\phi_n} \begin{pmatrix} x \\ y \end{pmatrix} = \begin{pmatrix} -r \sin(\phi_n) \\ r \cos(\phi_n) \end{pmatrix},
\end{align}
where $\vec{S}_n$ is the 2D vector field representation of $S^1$ for the $n$th level torus. Repeating the above process for $\vec{M}_x = 0, \nabla_x \times J_{\mathrm{tot}}$ and $\vec{M}_y = 0, \nabla_y \times J_{\mathrm{tot}}$ and combining gives three equations. For $\hat{z}$, the minimal, component-wise normalised form is
\begin{align} \label{eq:bigReqn}
   \begin{split}
       \nabla_y^2 - \nabla_x^2 - \nabla_z^2
       \;=\;
       &\left(\frac{4}{\pi^2}\,\Gamma_z\,\frac{\nabla \times \hat{\mathbf{z}}}{R_z^3}\right) - \frac{8}{R_z^2}
       \;-\;
       \left(\frac{4}{\pi^2}\,\Gamma_y\,\frac{\nabla \times \hat{\mathbf{y}}}{R_y^3}\right) + \frac{8}{R_y^2} \\
       &\;-\;
       \left(\frac{4}{\pi^2}\,\Gamma_x\,\frac{\nabla \times \hat{\mathbf{x}}}{R_x^3}\right) + \frac{8}{R_x^2}.
   \end{split}  
\end{align}

Continuing to evaluate the $z$ component note that the orthogonal scalar radii $R_x, R_y, R_z$ are not fixed. Straight-line impulse on the velocity field in the $x$ and $y$ direction occurs at the limit $R_z \rightarrow \infty$. Assessing all straight-line velocity perturbations in the $z$ component of \eqref{eq:bigReqn},
\begin{align} \label{eq:linear_pert1}
    \frac{4}{\pi^2}\left(\Gamma_y\frac{\nabla \times \hat{\mathbf{y}}}{R_y^3} - \Gamma_x\frac{\nabla \times \hat{\mathbf{x}}}{R_x^3}\right) - \left(\frac{8}{R_y^2}-\frac{8}{R_x^2}\right) = \nabla_x^2 - \nabla_y^2,\quad \begin{cases}
        R_x = R_y \\
        R_z \rightarrow \infty
    \end{cases}\\ \label{eq:linear_pert2}
    \frac{4}{\pi^2}\left(\Gamma_y\frac{\nabla \times \hat{\mathbf{y}}}{R_y^3} - \Gamma_z\frac{\nabla \times \hat{\mathbf{z}}}{R_z^3}\right) - \left(\frac{8}{R_y^2}-\frac{8}{R_z^2}\right) = \nabla_y^2 - \nabla_z^2,\quad \begin{cases}
        R_y = R_z \\
        R_x \rightarrow \infty
    \end{cases}\\ \label{eq:linear_pert3}
    \frac{4}{\pi^2}\left(\Gamma_z\frac{\nabla \times \hat{\mathbf{z}}}{R_z^3} - \Gamma_x\frac{\nabla \times \hat{\mathbf{x}}}{R_x^3}\right) - \left(\frac{8}{R_z^2}-\frac{8}{R_x^2}\right) = \nabla_z^2 - \nabla_x^2,\quad \begin{cases}
        R_z = R_x \\
        R_y \rightarrow \infty
    \end{cases}
\end{align}

Intuitively, select one radius $R_b\in\{R_x,R_y,R_z\}$ as the free scale and hold the other two fixed. This lets us solve the local angular momentum transfer over all directions ($4\pi$). A nearly straight-line perturbation along $\hat b$ is represented by $R_b\to\infty$ (equivalently $R_b\gg 1$).

With this choice, the three relations \eqref{eq:linear_pert1}--\eqref{eq:linear_pert3} follow by taking $R_b\to\infty$ in turn (for $b=z,y,x$) and equating the corresponding Laplacian differences. Cycling the free axis spans the momentum spectrum and yields the triplet
\begin{align}
\label{eq:ans_xy}
\frac{4}{\pi^2}\Big(\Gamma_x\frac{\nabla \times \hat{\mathbf{x}}}{R_x^3} - \Gamma_y\frac{\nabla \times \hat{\mathbf{y}}}{R_y^3}\Big) - \Big(\tfrac{8}{R_x^2}-\tfrac{8}{R_y^2}\Big) &= -\,\lap, \\[4pt]
\label{eq:ans_yz}
\frac{4}{\pi^2}\Big(\Gamma_y\frac{\nabla \times \hat{\mathbf{y}}}{R_y^3} - \Gamma_z\frac{\nabla \times \hat{\mathbf{z}}}{R_z^3}\Big) - \Big(\tfrac{8}{R_y^2}-\tfrac{8}{R_z^2}\Big) &= -\,\lap, \\[4pt]
\label{eq:ans_zx}
\frac{4}{\pi^2}\Big(\Gamma_z\frac{\nabla \times \hat{\mathbf{z}}}{R_z^3} - \Gamma_x\frac{\nabla \times \hat{\mathbf{x}}}{R_x^3}\Big) - \Big(\tfrac{8}{R_z^2}-\tfrac{8}{R_x^2}\Big) &= -\,\lap.
\end{align}
Together with \eqref{eq:current} and \eqref{eq:dep_pressure} this closes the construction in $\mathbb{R}^3$.

When 
\begin{align} \label{eq:paths}
    ds_x = ds_y,\qquad
    ds_y = ds_z,\qquad
    ds_z = ds_x,
\end{align}
we obtain a normalised, momentum-dependent Navier--Stokes--type model equation in the same algebraic form but with the component-wise closure entering explicitly:
\begin{align} \label{eq:non-dim-ans}
    \frac{\pt p_j}{p_j+4\pi \vec{u}_j} + \frac{4}{\pi^2}\,\Gamma_j\,\frac{\nabla \times \hat{\mathbf{j}}}{R^3_j} = \frac{8}{R_j^2} - \lap.
\end{align}

\begin{figure}
    \centering
    \includegraphics[width=1\linewidth]{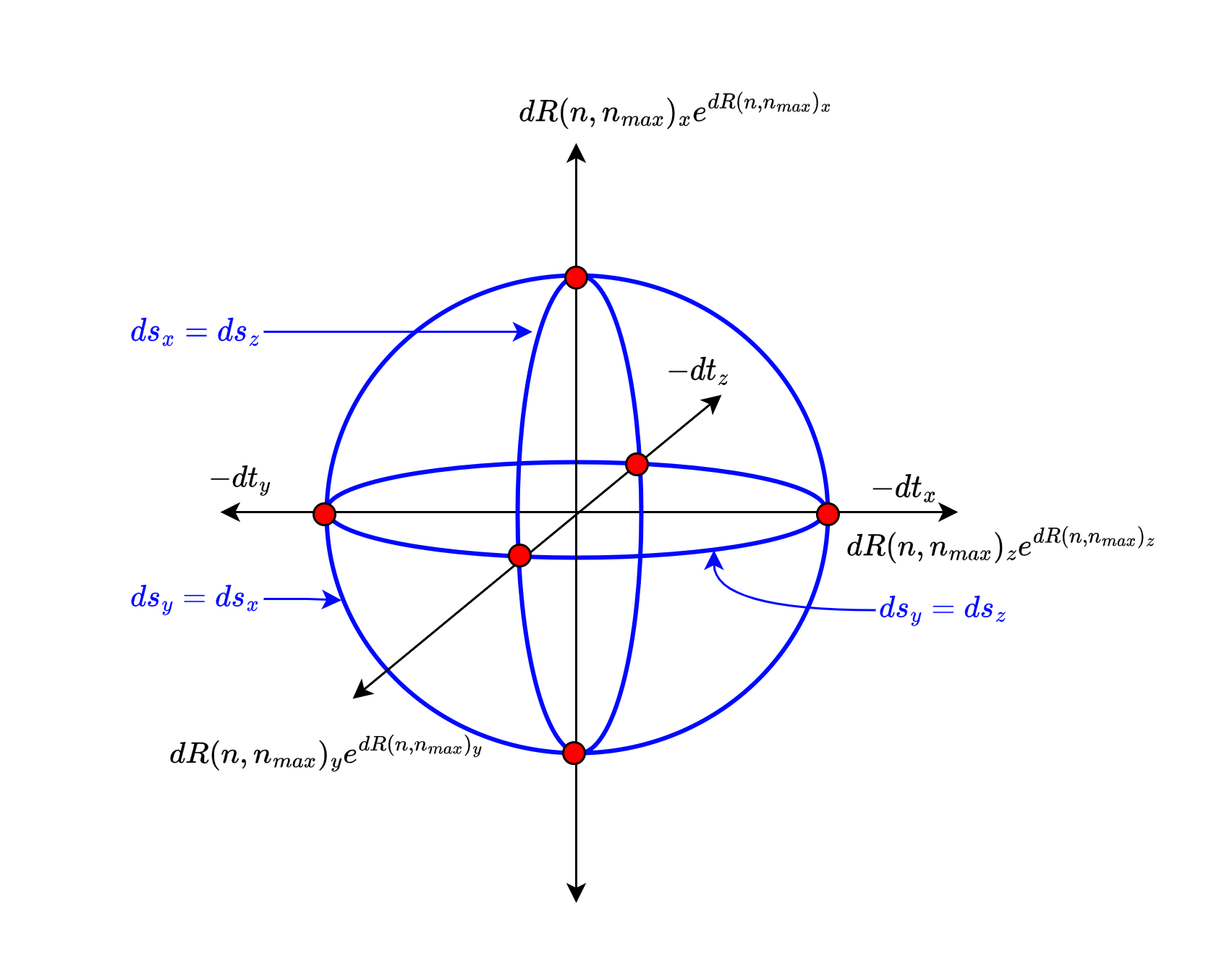}
    \caption{The $z$-oriented solutions of toroidal spin-oscillator spaces (red points) exist on hypersurfaces $dR_z e^{dR_z}$ that co-exist with 0, 1 or 2 independent time dimensions. We may interpret the opposing time directions associated with the $z$-hypersurface as an energy cascade balancing mechanism, or simply equal angular momentum dissipation rate into orthogonal directions, in this case $\hat{x}$ and $\hat{y}$. There are three positive time solutions and two time-independent solutions.}
    \label{fig:ds_soln}
\end{figure}

Since the right-hand side equals $-\lap$ in \eqref{eq:ans_xy}--\eqref{eq:ans_zx}, either a completely static solution exists, or the incompressibility condition holds,
\begin{align} \label{eq:div_eq}
    \nabla\cdot \vec{u} = 0.
\end{align}
The roots to the cubic \eqref{eq:non-dim-ans} provide the set of solutions for the Navier--Stokes--type model and may coincide with the fugacity of the system \cite{lee_polylogarithms_2009}. Shown in Fig.~\ref{fig:ds_soln} the intersection of orthogonal de Sitter paths at $n_{\max}$ allow for the solutions at the viscous limit, that is the local degeneracy of orbits and in the jerk frame. The interpretation of $p_j$ is the electric field flux in the direction $j \in \{x,y,z\}$ resulting from the momentum-dependent Coulomb interaction. The microscopic electric field is treated as a proxy for microscopic pressure gradients; macroscopic Ohmic-response language is not invoked here since the relevant quantities are not bulk conduction laws.
\\[4pt]
Setting $\vec{u} = 0$ implies a limit to the pulsation magnitude and that it be proportional to the wavevector as per the method in \cite{kolmogorov_local_1991}. Rewriting \eqref{eq:non-dim-ans} with this condition,
\begin{align} \label{eq:ns-fracs}
    \frac{4}{\pi^2}\,\Gamma_x\,\frac{\nabla \times \hat{\mathbf{x}}}{R_x^3} = \frac{8\,p_x}{R^2_x} - \lap p_x - \pt p_x ,\\
        \frac{4}{\pi^2}\,\Gamma_y\,\frac{\nabla \times \hat{\mathbf{y}}}{R_y^3} = \frac{8\,p_y}{R^2_y} - \lap p_y - \pt p_y ,\\
            \frac{4}{\pi^2}\,\Gamma_z\,\frac{\nabla \times \hat{\mathbf{z}}}{R_z^3} = \frac{8\,p_z}{R^2_z} - \lap p_z - \pt p_z ,
\end{align}
where 
\begin{align} \label{eq:v-diss-rate}
    \frac{4}{\pi^2}\left(\Gamma_x\frac{\nabla \times \hat{\mathbf{x}}}{R_x^3} + \Gamma_y\frac{\nabla \times \hat{\mathbf{y}}}{R_y^3}\right) = \frac{32\pi c \,\pt \mathbf{T}_z}{V^2}
\end{align} 
is the $z$-oriented volumetric dissipation rate with units $ \big[ \tfrac{C}{m^2 s} \tfrac{1}{m^3}\big]$ due to only the bound charge rotation and neglecting the contribution of electric field flux, differentiated by the use of $\mathbf{T}$ versus $\vec{T}^{(m)}$. Again using the intersection equalities from \eqref{eq:linear_pert1}--\eqref{eq:linear_pert3} and \eqref{eq:v-diss-rate}, we can rewrite \eqref{eq:ns-fracs} in volumetric dissipation terms oriented in $x, y$ and $z$. For $z$  
\begin{align} \label{eq:rfindroots}
    \nabla^{2}(p_x-p_y) = \frac{8(p_x-p_y)}{R^2} - \frac{32\pi}{R^3}\Big(\frac{\pt\mathbf{T}_z}{V^2} \Big), \quad c =1,\ R_x = R_y = R .
\end{align} 
Further we impose the condition 
\begin{align}
    \nabla^{2} (p_x - p_y) = 1,
\end{align}
which implies the jerk frame at the viscous heat-death limit, that is as cascade energy is constantly dumped into the spin-oscillator system the flux-derived angular momentum must change. In contrast to \cite{kolmogorov_local_1991} this is a non-equilibrium condition. Completing the above process along $x$ and $y$ and choosing roots of $R$ in the positive time direction (see Fig.~\ref{fig:ds_soln}), we can construct the energy cascade scale from the viscous heat death condition. 

\paragraph{Dissipative sub-range closure.}
At the dissipation boundary, after inertial transfer has saturated, the
residual reactive--dissipative response can be captured by an effective
higher-order closure; here this closure takes the form of a sixth-order
dissipative operator.

\[
\mathcal{D}=\nu_{6}\,\nabla^{6},\qquad \nu_{6}=\alpha_T\,T_b,
\]
with $\nu_{6}$ set by the microscopic magnetic toroidal susceptibility tensor $T_b$ and a numerical factor $\alpha_T$. The spectral dissipation density is therefore
\[
\frac{d\varepsilon}{dk}(k)=2\,\nu_{6}\,k^{6}\,E(k).
\]
If $d\varepsilon/dk$ varies slowly across this sub-range then the one-dimensional velocity spectrum obeys
\[
E(k)\ \propto\ \frac{1}{T_b}\,k^{-6}.
\]
This is analogous to Kolmogorov's identification of a dissipation weight where the spectrum is inversely proportional to the operator symbol.

\paragraph{Dissipation-range phenomenology and diagnostics.}
Classical closures and many experiments favour a \emph{stretched-exponential} falloff in the dissipation range, often written in the form
\[
E(k)\;\approx\; C\,\varepsilon^{2/3} k^{-5/3}\,\exp\!\big[-\beta\,(k\eta)^{m}\big],\qquad m\in[1,\,4/3],
\]
with representative constructions due to Pao and Pope and context from DNS surveys (e.g.\ Ishihara--Gotoh--Kaneda). 
First, from the degeneracy assumption, the longitudinal dissipation rate from \eqref{eq:dispersn} is
\begin{align}
    \frac{d\vec{D}}{dt} = \frac{k^{\frac{1+\sqrt{5}}{4}}}{4\pi}.
\end{align}
Using only the $t_+$ roots of $R$ in \eqref{eq:rfindroots} the de Sitter path interval of a closed loop about $b = \{\hat{x},\hat{y},\hat{z} \} \in \mathbb{R}^3$ becomes
\begin{align} \label{eq:ds_solutn}
    ds_+ = \Big( 16\pi^2\, dk^{-2\big({\frac{1+\sqrt{5}}{4}}\big)} + dR_b(k,r_{\min})^2\, e^{2\,dR_b(k,r_{\min})} \Big)^{\tfrac{1}{2}},
\end{align}
where $R_b = \big( \mathrm{Im}(R_1) + \mathrm{Re}(R_2) + \mathrm{Re}(R_3) \big)^{\tfrac{1}{2}}$, the applicable root components are
\begin{multline} \label{eq:roots}
\mathrm{Im}(R_1) = 2\, \mathrm{Im}\Bigg( \frac{2^{1/3} \big( \sqrt{3} \sqrt{27 \pi^2 p_{ab}^4 - 2 p_{ab}^3} - 9 \pi p_{ab}^2 \big)^{1/3}}{3^{2/3} p_{ab}} \\+ \frac{2^{2/3}}{3^{1/3} \big( \sqrt{3} \sqrt{27 \pi^2 p_{ab}^4 - 2 p_{ab}^3} - 9 \pi p_{ab}^2 \big)^{1/3}} \Bigg) ,
\end{multline}
\begin{multline*}
\mathrm{Re}(R_2) = -\, \mathrm{Re} \Bigg( \frac{2^{1/3} (1 - i \sqrt{3}) \big(\sqrt{3} \sqrt{27 \pi^2 p_{ab}^4 - 2 p_{ab}^3} - 9 \pi p_{ab}^2\big)^{1/3}}{3^{2/3} p_{ab}} \\- \frac{2^{2/3} (1 + i \sqrt{3})}{3^{1/3} \big(\sqrt{3} \sqrt{27 \pi^2 p_{ab}^4 - 2 p_{ab}^3} - 9 \pi p_{ab}^2\big)^{1/3}} \Bigg) , 
\end{multline*}
\begin{multline*}
\mathrm{Re}(R_3) = -\, \mathrm{Re} \Bigg(\frac{2^{1/3} (1 + i \sqrt{3}) \big(\sqrt{3} \sqrt{27 \pi^2 p_{ab}^4 - 2 p_{ab}^3} - 9 \pi p_{ab}^2\big)^{1/3}}{3^{2/3} p_{ab}} \\- \frac{2^{2/3} (1 + i \sqrt{3})}{3^{1/3} \big(\sqrt{3} \sqrt{27 \pi^2 p_{ab}^4 - 2 p_{ab}^3} - 9 \pi p_{ab}^2\big)^{1/3}} \Bigg) ,
\end{multline*}
where $ p_{ab} \in \{ (p_x-p_y), (p_z-p_x), (p_y-p_z) \}$.

\begin{figure}
    \centering
    \includegraphics[width=0.75\linewidth]{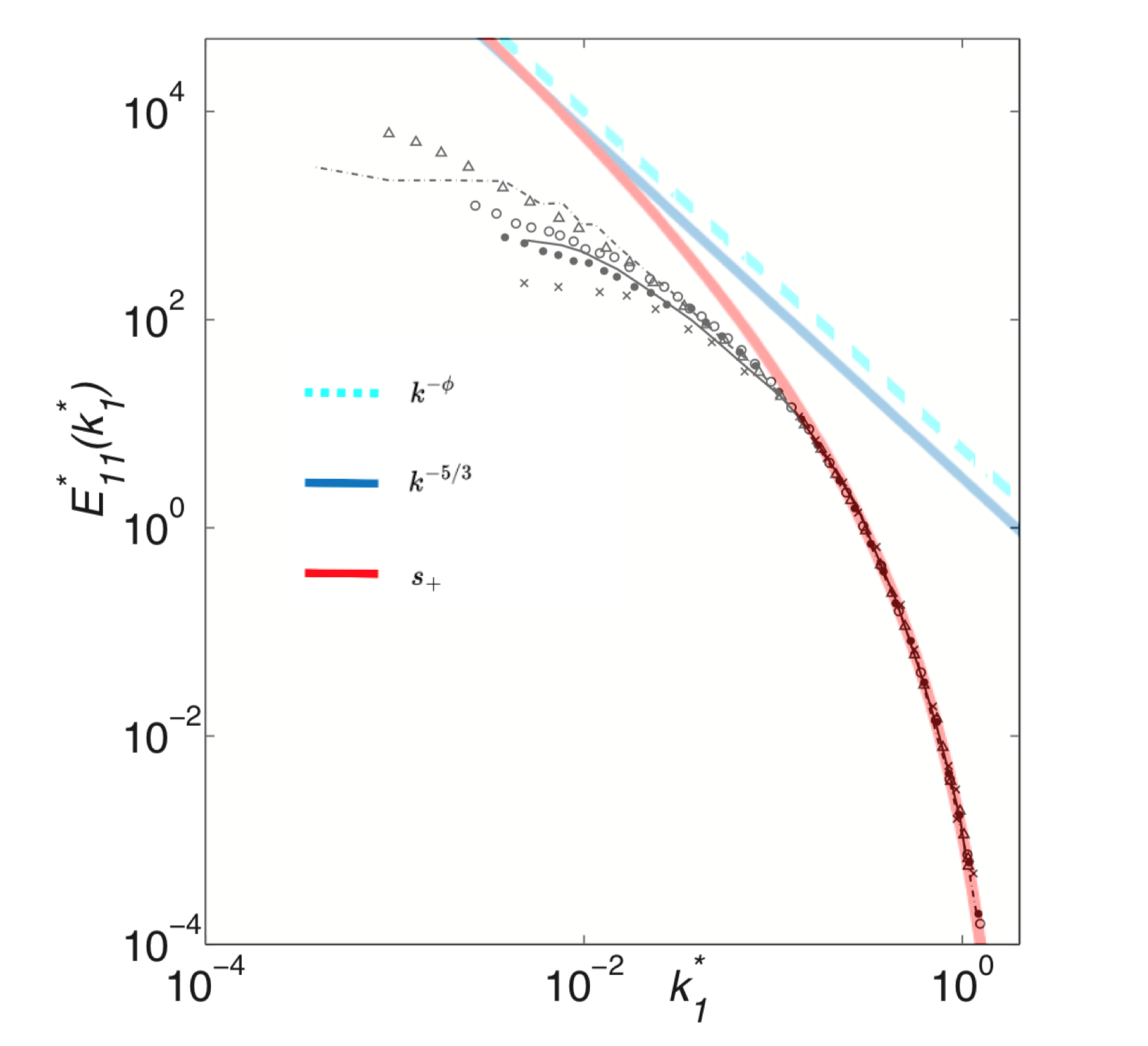}
    \caption{Red: solution of this work given by eq (\ref{eq:ds_solutn}) and (\ref{eq:roots}). Blue: K41 inertial scaling regime. Dashed Blue: non-viscous limit of eq(\ref{eq:ds_solutn}).  Normalised longitudinal spectra for various turbulence regimes and corresponding Taylor microscale Reynolds numbers $\mathrm{Re}_{\lambda}$. (x: $\mathrm{Re}_{\lambda}=40$, dots: 67, solid line: 65, circles: 89, triangles: 139, dash-dot line: 130). Fitting the dissipative curve to the turbulence data provides qualitative insight into the longitudinal energy gap at various frequencies. Adapted from \cite{antonia_collapse_2014}.}
    \label{fig:antonia}
\end{figure}

\paragraph{Conclusion.}
Unlike the inertial range, momentum (and charge-mediated momentum) transfer
in the viscous limit is tractable: dissipation suppresses multiscale
instability and the governing balance becomes Laplacian-dominated. By
retaining near-field electrodynamic structure under a relaxed LWA and
introducing the dimensionless rate ratio
$\Gamma_j=\partial_t D'_j/\partial_t D_j$, we obtain a closed viscous-limit
Navier--Stokes--type model with $\Delta$ as the
smoothing operator. Under the invariant-tori degeneracy hypothesis the
anisotropy is fixed by $\gr/2$, and the dissipation-range spectrum follows
$E(k)\propto T_b^{-1}k^{-6}$ under a sixth-order closure.
The solution for the model with suppressed intermittency is given by \eqref{eq:ds_solutn} and \eqref{eq:roots}. $\Box$

\pagebreak
\printbibliography
\end{document}